\documentclass[%
 superscriptaddress,
 amsmath,amssymb,
 aps,
pra,
floatfix,
twocolumn
]{revtex4-2}

\usepackage{braket}
\usepackage{graphicx}% Include figure files
\usepackage{dcolumn}% Align table columns on decimal point
\usepackage{bm}% bold math
\usepackage{hyperref}% add hypertext capabilities
\usepackage{subfigure}
\usepackage{siunitx}
\usepackage{lipsum}
\usepackage{bbold}
\newcommand{\ie}{i.\,e.}
\newcommand{\eg}{e.\,g.}

\newcommand{\ped}[1]{_\text{#1}}

\newcommand{\npop}{N\ped{pop}}
\newcommand{\ngen}{N\ped{gen}}
\newcommand{\len}{L}
\newcommand{\pmut}{P\ped{m}}
\newcommand{\pcx}{P\ped{c}}
\newcommand{\pind}{P\ped{ind}}
\newcommand{\tournsize}{N\ped{T}}

\newcommand{\nspin}{n}
\newcommand{\n}{N}
\newcommand{\ham}{H}
\newcommand{\hamx}{\ham_x}
\newcommand{\hamz}{\ham_z}
\newcommand{\hamcd}{\ham\ped{od}}
\newcommand{\tf}{T}
\newcommand{\mingap}{\Delta}

\newcommand{\iu}{i}

\usepackage{xcolor}

\graphicspath{{pictures/}}

\begin{document}

\preprint{APS/123-QED}

\title{Genetic optimization of quantum annealing}

\author{Pratibha Raghupati Hegde}
\email{pratibharaghupati.hegde@unina.it}
\affiliation{Dipartimento di Fisica “E. Pancini”, Università di Napoli Federico II}

\author{Gianluca Passarelli}
\affiliation{CNR-SPIN, c/o Complesso di Monte S. Angelo, via Cinthia - 80126 - Napoli, Italy}
%\affiliation{Dipartimento di Fisica “E. Pancini”, Università di Napoli Federico II}
%\affiliation{CNR-SPIN, c/o Complesso di Monte S. Angelo, via Cinthia - 80126 - Napoli, Italy}
\author{Annarita Scocco}
\affiliation{Dipartimento di Fisica “E. Pancini”, Università di Napoli Federico II}
\author{Procolo Lucignano}
\affiliation{Dipartimento di Fisica “E. Pancini”, Università di Napoli Federico II}

%


\begin{abstract}
The study of optimal control of quantum annealing by modulating the pace of evolution and by introducing a counterdiabatic potential has gained significant attention in recent times. In this work, we present a numerical approach based on genetic algorithms to improve the performance of quantum annealing, which evades the Landau-Zener transitions to navigate to the ground state of the final Hamiltonian with high probability. We optimize the annealing schedules starting from polynomial ansatz by treating their coefficients as chromosomes of the genetic algorithm. We also explore shortcuts to adiabaticity by computing a practically feasible $k$-local optimal driving operator, showing that even for $k=1$ we achieve substantial improvement of the fidelity over the standard annealing solution.  With these genetically optimized annealing schedules and/or optimal driving operators, we are able to perform quantum annealing in relatively short time-scales and with larger fidelity compared to traditional approaches. 

\end{abstract}

\maketitle

%\tableofcontents

\section{\label{sec:intro}Introduction}

    Small spectral gaps are the bottleneck of adiabatic quantum computation and quantum annealing~\cite{kadowaki-nishimori:qa,santoro:qa,jorg:energy-gaps,knysh:bottlenecks}. In these paradigms of quantum computation, the goal is to read the ground state of a target Hamiltonian $ \hamz $, encoding an NP-complete or NP-hard problem~\cite{lucas:np-complete}. Starting from the (easy to prepare) ground state $ \ket{\psi(0)} $ of a transverse field Hamiltonian $ \hamx = -\Gamma \sum_{i=1}^\nspin \sigma^x_i $, where $ \nspin $ is the number of qubits and $ \Gamma $ is the strength of the transverse field, the system is evolved with time-dependent Hamiltonian $ \ham_0(t) = A(t) \hamx + B(t) \hamz $. The annealing schedule is given by the pair $ \set{A(t), B(t)} $, satisfying $ A(0) \gg B(0) $ and $ 0 = A(\tf) \ll B(\tf) $, where $ \tf $ is the annealing time.
	At $ t = \tf $, the system is found in the target ground state with a large probability, provided that $ \tf $ is longer than the inverse square of the smallest gap between the ground state and the first excited state~\cite{lidar:adiabatic-theorem}. During the dynamics the system may cross a quantum phase transition~\cite{sachdev:qpt}, correspondingly the gap takes its minimum value $ \mingap = \min_t [E_1(t) - E_0(t)] $ which results in long  annealing times to satisfy the adiabatic condition, thus  making the algorithm ineffective.

	If the annealing time $\tf$ is shorter than what predicted by the adiabatic theorem, the fidelity of the final solution is compromised and, if $\tf$ is longer, the system suffers decoherence. Therefore the goal, here, is modifying the annealing dynamics in order to achieve high fidelities even breaking the adiabatic criterion, before decoherence sets in.

	This can be achieved taking benefit of different improved schemes. We here mention optimal control theory \cite{Koch:EPJD2015} which  is limited, in principle, only by the quantum speed limit \cite{Caneva:PRL2009,Hegerfeldt:PRL2013}, or shortcuts to adiabaticity (STA)~ \cite{torrentegui:sta,delcampo:sta,fazio:sta,fazio:sta-local,campbell:sta,obinna:sta,delcampo:sta-work-fluctuations,yehong:st-by-substitute, santos:sta, santos:sta2, santos:sta3, santos:sta4, santos:sta5} or modulating in a controlled way the annealing schedules ~\cite{susa-nishimori-timesch-optimization,matsuura-timesch-optimization,bolte-genetic-timesch-optimizing,Roland:time-sch}.

%	Modifying the annealing schedules $A(t)$, and $B(t)$.  However, one can control the rate of evolution in between $0$ and $\tf$, to improve the performance of computation. Therefore, a study of optimized choices of annealing schedules is crucial in this field~\cite{susa-nishimori-timesch-optimization,matsuura-timesch-optimization,bolte-genetic-timesch-optimizing}. 
	 
A possible STA consists in adopting counterdiabatic (CD) driving~\cite{torrentegui:sta,delcampo:sta,fazio:sta,fazio:sta-local,campbell:sta,obinna:sta,delcampo:sta-work-fluctuations,yehong:st-by-substitute}. In transitionless or CD driving, a time-dependent potential $ H_\text{cd}(t) $ is added to the unperturbed Hamiltonian $ \ham_0(t) $ so that diabatic Landau-Zener transitions are completely suppressed at all times and for all choices of the annealing time $ \tf $. The total Hamiltonian reads $ \ham(t) = \ham_0(t) + H_\text{cd}(t) $. The CD operator satisfies the constraint $ H_\text{cd}(0) = H_\text{cd}(\tf) = 0 $ and does not modify the starting and target Hamiltonians. Computing the exact CD potential requires the knowledge of the (generally unknown) instantaneous spectrum of the Hamiltonian $ \ham_0(t) $. Moreover, the resulting operator is highly nonlocal, hardly implementable on actual quantum machines, and generally unbounded in the thermodynamic limit~\cite{berry:exact-cd}. 
		
Recently, a lot of effort has been put to building approximate CD potentials. In some very simple cases, such as the Ising model with longitudinal and transverse fields, linear combination of local operators provide good approximations of the CD potential, \eg, $ H_\text{cd}(t) \approx \sum_k \alpha_k(t) O_k $. The $ O_k $ operators are generally Hermitian products of a small number of Pauli operators. The coefficients $ \alpha_k(t) $ can be determined by variational optimization~\cite{polkovnikov:local-cd, hartmann2019:local-cd}. For more complicated many-body Hamiltonians, other choices for operators $ O_k $ involve nested commutators between $ \ham_0(t) $ and its time derivative~\cite{polkovnikov:nested-cd}. However in the former case, we do not know in advance which and how many local operators are needed to build  a good ``enough'' CD operator. In the latter case, nested commutators can be highly non-local, as much as the exact CD potential. Moreover, the number of nested commutators is expected to diverge in the thermodynamic limit when the system undergoes a quantum phase transition~\cite{passarelli:PRR2020}. 
		
In this paper, we derive an alternative route and we focus on the study of optimal annealing schedules $A(t),B(t)$ as well as on an optimal driving (OD) potential $ \hamcd(t) $  that are variationally improved so to achieve the maximum fidelity at the final time $\tf$. The search for  variational minima is approached by computational intelligence tools \cite{bolte-genetic-timesch-optimizing}, in particular we adopt  a genetic algorithm, \ie, an evolutionary strategy inspired by the Darwinian theory of the survival of the fittest~\cite{YAO1993707}.  We consider time schedules that are polynomial functions of time and we consider local operators for the OD. In our approach, the coefficients of the polynomials and the OD operator are represented as a real-valued chromosome. Each chromosome is characterized by a fitness value. At each generation, chromosomes will mate and randomly mutate. Only the fittest individuals will survive to the next generation. 
We show that a simple choice of the fitness function can lead to optimized annealing schedules as well as to OD potentials that largely improve the fidelity of the target quantum ground state of $ \hamz $, compared to the bare case. We discuss the adiabatic quantum computation of a prototypical system, the ferromagnetic $ p $-spin model, an exactly solvable model with a  nontrivial phase diagram, which  encodes a Grover-like search \cite{grover1996,Roland:time-sch} for large and odd $p$.
    
    This paper is organized as follows. In Sec.~\ref{sec:problem}, we describe the ferromagnetic $p$-spin model. In Sec.~\ref{sec:methods}, we introduce the genetic algorithms and the construction of chromosomes for the problems of optimization of annealing schedules and OD potentials. We also define fitness functions for single objective genetic algorithm (SOGAs) and multi-objective genetic algorithms (MOGAs). In Sec.~\ref{sec:results}, we present the results obtained by optimizing the annealing schedules, OD potentials individually and together using genetic algorithms. In Sec.~\ref{sec:ising} we discuss the possibility of extending our techniques to the quantum annealing of random Ising models. We finally derive our conclusions in Sec.~\ref{sec:conclusions}.
    \begin{figure*}
		\centering
		\includegraphics[width=\textwidth]{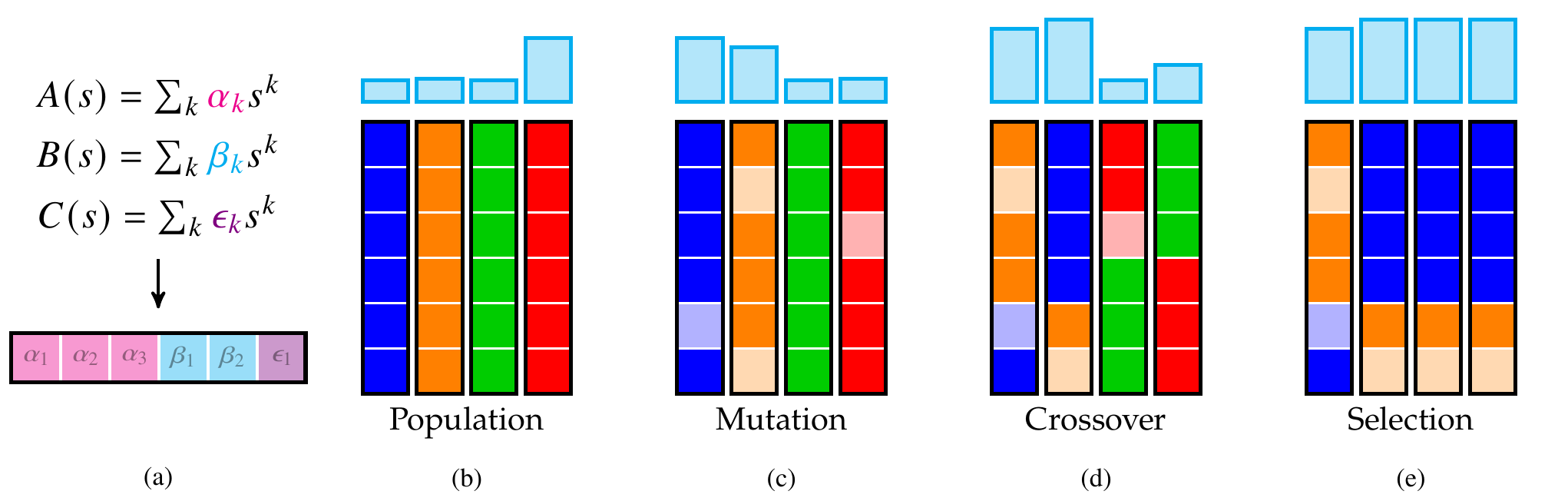}
		\caption{Cartoon of our genetic algorithm. a) The free parameters of the annealing schedules are stored into a chromosome. b) We first randomly generate $\npop$ individuals. c) Then random gene mutation occurs in each individual. d) Then we apply two individual crossover. e) We select the fittest individuals and start again from c) until convergence. The azure bars identify the fitness values: the larger the better.}
		\label{fig:genetic-sketch}
    \end{figure*}

\section{\label{sec:problem}Problem definition}
    In this paper, we consider the fully-connected ferromagnetic $ p $-spin model~\cite{derrida:p-spin,gross:p-spin} as case study. The Hamiltonian of this model is
    \begin{equation}
    \hamz = -J\nspin {\left(\frac{1}{\nspin}\sum_{i = 1}^\nspin \sigma^z_i\right)}^p,
    \label{ham:p-spin}
    \end{equation}
    with $ J > 0 $ and $ p \ge 2 $. For odd $ p $, its ground state is ferromagnetic with all qubits in the state $ \ket{0} $. For even $ p $, the ground state manifold is 2-dimensional ($ \ket{00\cdots0} $ and $ \ket{11\cdots1} $) due to the $ Z_2 $ symmetry. If we study the quantum annealing with time-dependent Hamiltonian $ \ham_0(t) $ using as a target Hamiltonian $\hamz$ defined in  Eq.~\eqref{ham:p-spin}, we observe  a dynamical quantum phase transition separating a paramagnetic phase (at short times) from a ferromagnetic phase (at long times). For $ p = 2 $, the QPT is of second order, while for $ p \ge 3 $ it is of first order. The latter is the hardest case for quantum annealing, as the minimal gap $ \mingap $ closes exponentially as a function of $ \nspin $~\cite{bapst:p-spin}. This feature motivates the broad interest in this system as a toy model of NP-hard problems~\cite{nishimori:p-spin-1,nishimori:p-spin-2,nishimori:p-spin-3, nishimori:p-spin-4,nishimori:p-spin-5,nishimori:p-spin-6,nishimori:p-spin-7,passarelli:p-spin,passarelli:pausing,acampora:genetic, passarelli:PRR2020,passarelli:reverse-pspin}.
	
	The model Hamiltonian is permutationally invariant and commutes with the total spin operator $ S^2 $ at all times. The starting and the target state belong to the subspace with maximum spin $ S = \nspin / 2 $ and the dynamics will occur within the same (maximum spin) subspace. Therefore, we can work in this $ (\n = \nspin + 1) $-dimensional sector. In the following, we will consider $ J $ as unit of energy. Times are expressed in units of $ J^{-1} $ ($ \hslash = 1 $ here and in the following).

    We perform adiabatic evolutions of the system described by the $ p $-spin model assisted by genetic algorithms. We aim at improving the final state fidelity of the system by following three strategies: a) optimizing annealing schedules b) optimizing local OD with the traditional linear annealing schedules and c) optimizing both annealing schedules and local OD operator. These strategies are explained in detail later in the paper, see Sec.~\ref{sec:methods}. Further, we choose an annealing time sufficiently shorter than the timescale $T_\text{AD}$ predicted by the adiabatic theorem, \ie,
    \begin{equation}\label{eq:adiabatic-timescale}
        T_\text{AD}= \max_{t \in [0, T]}\frac{\lvert \langle \epsilon_0(t)\vert \partial_t H(t) \vert \epsilon_1(t)\rangle\rvert}{\lvert \epsilon_1(t) -\epsilon_0(t) \rvert ^2}.
    \end{equation}
     
\section{\label{sec:methods} Methods: Genetic algorithms}
    We use a class of evolutionary algorithms known as genetic algorithms to find optimized annealing schedules for adiabatic evolutions. In addition, we also manage to demonstrate the efficiency of genetic algorithms in the paradigm of shortcuts to adiabaticity by finding optimized, local OD operators. 
    %Further, we compute the time dependent function $C(s)$ which dictates the CD evolution using genetic optimization.
    
    Genetic algorithms are inspired by Darwin's theory of evolution. These algorithms offer solutions to optimization problems conditioned by a single objective or multiple objectives~\cite{acampora:genetic,acampora:genetic2,deb:nsga2,deap}. In both cases, the possible solutions to the problem are encoded as a string of real numbers called chromosomes. The construction of a chromosome depends on the optimization problem. In this article, broadly speaking, we address three optimization problems, all of which aid the performance of adiabatic evolution, i.\,e., finding the system in a ground state of the problem Hamiltonian $\hamz$ with maximum probability at the end of the evolution. The three problems are as follows.
    
    \subsection{Optimization of annealing schedules}\label{sec:optimization-annealsch} 
        Here we try to optimize the performance of quantum annealing by optimizing its annealing schedules $A(t)$ and $B(t)$~\cite{matsuura-timesch-optimization, bolte-genetic-timesch-optimizing, susa-nishimori-timesch-optimization}. Firstly, we express the annealing schedules as dimensionless time functions of $s=t/T$ throughout this paper. We consider polynomial expansions of $A(s)$ and $B(s)$ as candidates for the possible annealing time schedules, \ie, $A(s, \alpha) = \sum_{i=1}^{k_a+1} \alpha_i s^{i}$, $B(s, \beta) = \sum_{j=1}^{k_b+1} \beta_j s^{j}$. Moreover, these time-dependent functions have to satisfy the boundary conditions, $A(0)=1,\,A(1)=0$ and $B(0)=0,\, B(1)=1$, and therefore can expressed as
        \begin{equation}\label{eq:annealing-sch}
          \begin{aligned}
            A(s, \alpha) &= 1+ \sum\limits_{i=1}^{k_a} \alpha_{i} s^{i} + \left(-1-\sum\limits_{i=1} ^{k_a} \alpha_i\right)s^{k_a+1}, \\
            B(s, \beta) &= \sum\limits_{j=1}^{k_b} \beta_{j} s^{j} + \left(1-\sum\limits_{j=1} ^{k_b} \beta_j\right)s^{k_b+1}.
        \end{aligned}  
        \end{equation}
        
        We optimize the coefficients of these polynomial expansions as chromosomes of the genetic algorithm and  the structure of the chromosome for this problem is
        \begin{equation}\label{eq:chrom-annealing-sch}
            D_1 = \left[\alpha_1, \alpha_2, \cdots \alpha_{k_a}, \beta_1, \beta_2, \cdots \beta_{k_b} \right].
        \end{equation}
        The length of the chromosome is $k_a + k_b$. 
        
    \subsection{Optimization of the local OD operator}\label{sec:optimization-cd}
        In this section, we adopt the strategy of shortcuts to adiabaticity to optimize the performance of quantum annealing~\cite{berry:exact-cd, polkovnikov:local-cd, polkovnikov:nested-cd, passarelli:PRR2020, hartmann2019:local-cd}. Keeping the annealing schedules to be fixed and as linear functions, \ie, $A(s) = 1-s$ and $ B(s)=s$, we optimize an OD operator which successfully avoids Landau-Zener transitions resulting in a better fidelity of the state of the system with the exact ground state at $ t = T $. We assume that the OD operator $\hamcd(s)$ can be expanded as the sum of local spin operators,
        \begin{equation}\label{eq:local-cd-op}
            \hamcd(s, \gamma)=C(s)\sum\limits_{i=1}^d \gamma_i O_i,
        \end{equation}
        where $O_i$ are the total spins along the $x$, $y$ and $z$ directions, \ie, $S_x$, $S_y$ and $S_z$, and their products. Especially, we consider only single local operators and cumulatively add the set of all possible 2-spin operators and the set of all 3-spin operators. These local operators can be explicitly written as
        \begin{equation}\label{eq:cd-operators}
        \begin{aligned}
             H_\text{od}(s, \gamma)_{d=3} &= C(s)\sum\limits_{i=1}^3 \gamma_i S_i,\\
             H_\text{od}(s, \gamma)_{d=9} &=  H_\text{od}(s)_{d=3}+C(s)\sum_{i,j=1}^3\gamma_{i,j} S_i S_j,\\
             H_\text{od}(s, \gamma)_{d=21} &= H_\text{od}(s)_{d=9}+C(s) \sum_{i,j,k=1}^3 \gamma_{i, j, k} S_i S_j S_k.
        \end{aligned}
        \end{equation}
        
          The chromosome of the genetic genetic algorithm for this problem is the set of coefficients of the local operators, 
        \begin{equation}\label{eq:chrom-local-cd-op}
            D_2 = \left[\gamma_1, \cdots , \gamma_3, \gamma_{11}, \gamma_{12}, \cdots, \gamma_{33}, \gamma_{111}, \gamma_{112}, \cdots, \gamma_{333} \right],
        \end{equation}
%        \pcom{  The notation is not very clear. The elements of $D_2$ are labeled with  just a single index $\gamma_j$ while those of the OD Hamiltonian can have two or three indexes }
        whose length is equal to the number of local operators, $d$. In this work, we are able to achieve remarkable results by optimizing the local OD operator with only single spin operators, \ie, $H_\text{od}(s, \gamma)_{d=3} = C(s)\sum_{i=1}^3 \gamma_i S_i$, and therefore we discuss and demonstrate our results for the case with $d=3$. The higher terms of 2-spin and 3-spin operators are omitted since they do not produce any significant improvements. The time schedule $C(s)$ is fixed in this approach and is given by $C(s)=A(s)B(s)=(1-s)s$. The function $C(s)$ controls the pace of evolution dictated by the OD operator $\hamcd(s)$.
        
    \subsection{Optimization of the time schedules and the local OD}\label{sec:optimization-all}
        Finally, here we optimize the annealing schedules $A(s)$, $B(s)$ and the local OD operator altogether~\cite{susa-nishimori-timesch-optimization, matsuura-timesch-optimization}. The time schedule $C(s)$  is optimized by absorbing it  as the coefficients of the local OD operators, \ie,  $\hamcd(s) = \sum_{i=1}^d C_i(s) O_i$. We consider each $C_i(s)$ to be a polynomial of order $k_c+1$, which satisfies the boundary condition $C_i(0) = 0$ and $C_i(1)=0$. Therefore, the OD operator can be explicitly written as
        \begin{equation}\label{eq:cd-ts}
            \hamcd(s, \epsilon) = \sum\limits_{i=1}^d\left(\sum\limits_{j=1}^{k_c} \epsilon_{ji} s^{j} + \left(-\sum\limits_{j=1} ^{k_c} \epsilon_{ji}\right)s^{k_c+1}\right) O_i.
        \end{equation}
        We optimize the free parameters $\epsilon_{ji}$, in addition to the free parameters $\alpha_i$ and $\beta_i$ in Eq.~\eqref{eq:chrom-annealing-sch}. The time-dependent Hamiltonian of the system is given by
        \begin{equation}\label{eq:var-total-hamiltonian}
            H(s) = A(s, \alpha)\hamx + B(s, \beta)\hamz + \hamcd(s, \epsilon),
        \end{equation}
        where 
        \begin{equation*}
        \begin{aligned}
         \alpha &= \{\alpha_1,...,\alpha_{k_a}\}\\
         \beta &= \{\beta_1,...,\beta_{k_b}\}\\
         \epsilon &= \{\epsilon_{11},...,\epsilon_{k_{c}d}\}.\\
         \end{aligned}
        \end{equation*}
        Therefore, the chromosome for this optimization problem can be expressed as
        \begin{equation}\label{eq:chrom-cd-ts}
            D_3 = \left[\alpha_1, \cdots \alpha_{k_a}, \beta_1,\cdots, \beta_{k_b}, \epsilon_{11},  \cdots \epsilon_{k_{c} 1},\cdots \epsilon_{k_{c} d} \right].
        \end{equation}
        The length of the chromosome in this case is $k_a+k_b+(d\times k_c)$. Again here, we are able to obtain large fidelity of the state of the system by considering only single spin operators in the expansion of the local OD operator. Therefore we stick to the case of $d=3$.
    
    %Having defined the problems and the associated chromosomes, in the remaining of this section, we describe the workflow of genetic algorithms. In this paper, we implement both single objective genetic algorithms(SOGA), as well as multi-objective genetic algorithms(MOGA).
    The key aspect of genetic algorithms is the definition of fitness function. It is a function which takes each chromosome as a variable and gives it a fitness value according to the quality of the solution generated by the given chromosome. In the course of a genetic algorithm, we intend to either maximize or minimize this fitness function. Depending on the number of conditions the chromosomes have to satisfy, the genetic algorithms are characterized by fitness functions which are single objective or multi-objective. In the remaining of this section, we describe the fitness function and the workflow of Single Objective Genetic Algorithms (SOGAs) and Multi-Objective Genetic Algorithms (MOGAs).
    
    \subsection{Single objective genetic algorithms}
    SOGAs follow the workflow of a standard genetic algorithm. We define the fitness of each chromosome as the fidelity, which is the ground state probability at $ t = \tf $, \ie,
	\begin{equation}\label{eq:fitness-function}
		f_\text{so} \equiv  P_\text{gs}(\tf)={\lvert \langle E_0(\tf) | U(\tf) | \psi(0) \rangle \rvert}^2 ,
	\end{equation}
	where $ U(t, 0) = \mathcal{T}_+ \exp\lbrace-\iu \int_{0}^{t} [\ham(t')] dt'\rbrace $ is the time evolution operator and $ \mathcal{T}_+ $ is the time ordering~\footnote{The time evolution is computed with the QuTiP toolbox~\cite{qutip1,qutip2}.}. An alternative fitness function would be to use the negative of the mean energy at the final time $\tf$, \ie, $-\langle \psi(0) | U^\dagger(\tf, 0) \hamz U(\tf, 0) | \psi(0)\rangle $. This choice does not require the knowledge of any spectral property of the Hamiltonian. The fittest individuals, maximizing $ f_\text{so} $, are those with larger fidelities and are likely to survive along generations. At the end of the genetic optimization, we will obtain a chromosome defined according to the problem. However, all three problems considered here aim at giving a higher fidelity.
   
    We initialize a starting population of $ \npop $ individuals, whose genes are randomly extracted in the interval $ [g\ped{min}, g\ped{max}] $. Then, we repeatedly apply the three genetic operators (mutation, crossover and selection~\cite{Herrera03ataxonomy}) until a convergent solution is achieved. The genetic algorithm is implemented by using the DEAP package~\cite{deap}. Here, we briefly describe the genetic operators adopted, also sketched in Fig.~\ref{fig:genetic-sketch}.
	
\textit{i) Gaussian mutation}---Among the population of individuals, random individuals are selected with a probability $ \pmut $ for mutation. Each gene is independently mutated with a probability $ \pind $, by adding a normal variable, extracted from a Gaussian with mean value $ \mu = 0 $ and variance $ \sigma^2 = 1 $ [see Fig.~\ref{fig:genetic-sketch}(c)]. The mutation probability of each gene, \ie, the product $ \pmut\pind $, should be neither too large nor too small (a quantitative description is given in Appendix~\ref{sec:hparameters}). In the former case, the genetic algorithm will turn into a random search. In the latter case, the algorithm would be nonergodic. These random mutations increase variability in the population and reduce the probability of being trapped in local minima.
	
\textit{ii) Two-point crossover}---After mutation process, we randomly select two parents from the chromosome population. Two random integers are randomly extracted in the interval $ [0, \len - 1] $, where $\len$ is the length of the chromosome, which is the number of free parameters to be optimized using a genetic algorithm and is problem-specific. Two children are produced by mixing alternating parts of the two parents, obtained by cutting the chromosomes at the two extracted indices [see Fig.~\ref{fig:genetic-sketch}(d)]. Note that the exchange of the fragments is only symbolic in Fig.~\ref{fig:genetic-sketch}(d) and represents a one-point crossover for the sake of visual clarity. In our experiments, we resort to a two-point crossover operator which yields the fastest convergence in this case. The whole process occurs with a probability $ \pcx $. Small $ \pcx $ ensures slow but accurate convergence to the optimal solution. On the other hand, large $ \pcx $ ensures quick convergence but can lead to sub optimal solutions. Hence, $ \pcx $ has to be carefully tuned to find a compromise between speed of convergence and accuracy of the solution.
	
	\textit{iii) Selection by tournament}---After mutation and crossover, a new population is produced. $ \tournsize $ competitors are selected from the population and their fitness is compared [see Fig.~\ref{fig:genetic-sketch}(e)]. Only the fittest individual survives to the next generation. This tournament is repeated until we obtain a new set of $ \npop $ individuals.

	\subsection{Multi-objective genetic algorithms}
	
	While SOGAs aim at maximizing the ground state probability at the final time $\tf$, they sometime lead to practically not feasible solutions during the time of evolution. For example, some of the solutions returned by the algorithm can have energy level crossings between the ground state and the first excited state. In an attempt to avoid these solutions produced by SOGAs, we add another objective to the fitness function.  Other than maximizing the fidelity at $ t = \tf $, we choose to maximize it together with the area under the curve of the instantaneous ground state probabilities of the system computed at $N_t$ time intervals. The latter assures that the ground state occupation is maximum at all the intermediate times, in the spirit of counterdiabatic dynamics. 
	The ground state probability at time $t$ is given by $P_\text{gs}(t) = {\lvert \langle E_0(t) | U(t) | \psi(0) \rangle \rvert}^2$. The fitness of a chromosome in MOGA are defined as
	\begin{equation}\label{eq:fitness-mo}
	    f_\text{mo} \equiv \left\{ \frac{1}{\tf}\int_0^\tf P_\text{gs}(t) \, dt,  P_\text{gs}(\tf) \right\}.
	 \end{equation}
	We stress here the fact that this is not the same as imposing local adiabaticity as by \textcite{Roland:time-sch}.
	MOGAs deviate from the standard genetic algorithms. In particular, they work using the strategy of Non-dominated Sorting Genetic Algorithm II (NSGA-II)~\cite{deb:nsga2, chivilikhin:nsga2}. NSGA-II uses an elitist method of evolutionary algorithms. The parent and offspring generations are clubbed together and are ranked into fronts based on non-dominated sorting. The population of the following generation is filled with the best fronts until $\npop$ is reached. In case that only some chromosomes have to be selected from a front in the process, the most diverse solutions are chosen based on the crowding distance. Given the new population, by the above non-dominated sorting process, the chromosomes undergo selection (a binary tournament selected based on both rank and crowding distance), mutation and crossover processes. In the end of $\ngen$  generations, the Pareto optimal front with the best ranking is obtained. The details of selecting the chromosome from the Pareto optimal front is given in Appendix~\ref{sec:pof}.
	\begin{figure}
        \centering
        \includegraphics[width=\columnwidth]{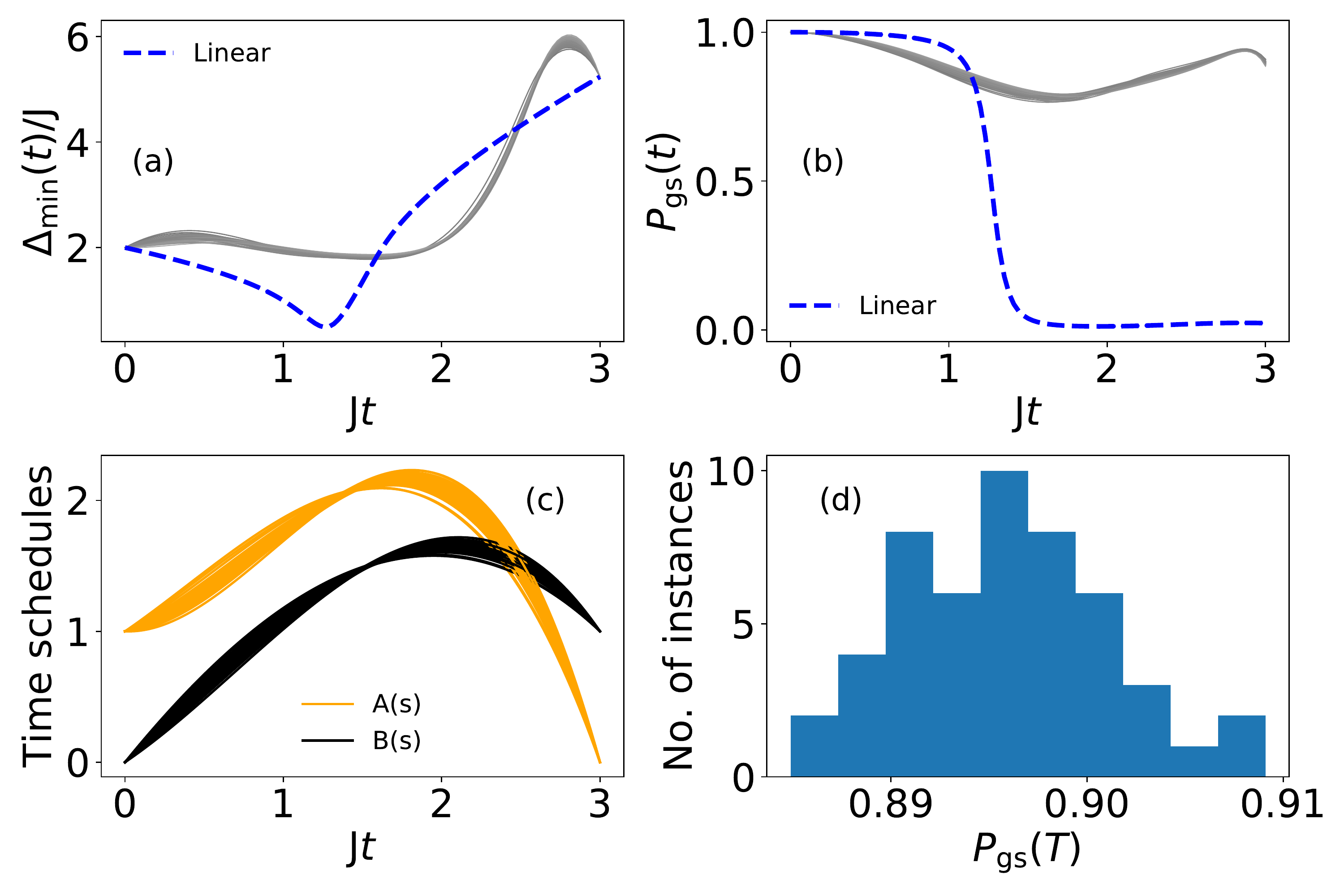}
        \caption{Summary of the results obtained by using the optimized annealing schedules for solving the $p$-spin model using SOGA. We investigate the system with $15$ spins with both the annealing schedules $A(s)$ and $B(s)$ expanded up to a degree of $3$. In other words, $k_a=k_b=2$. In panel (a), we show the instantaneous energy gaps during the dynamics of the adiabatic evolution, in panel (b) we show the instantaneous ground state probability using the optimised schedules and by using a simple linear schedule, in panel (c) we show the annealing schedules $A(s)$ and $B(s)$, and in panel (d) we show the histogram of the fidelities for 50 runs of the algorithm. }
        \label{fig:annealsch-so}
    \end{figure}
    \begin{figure}
        \centering
        \includegraphics[width=\columnwidth]{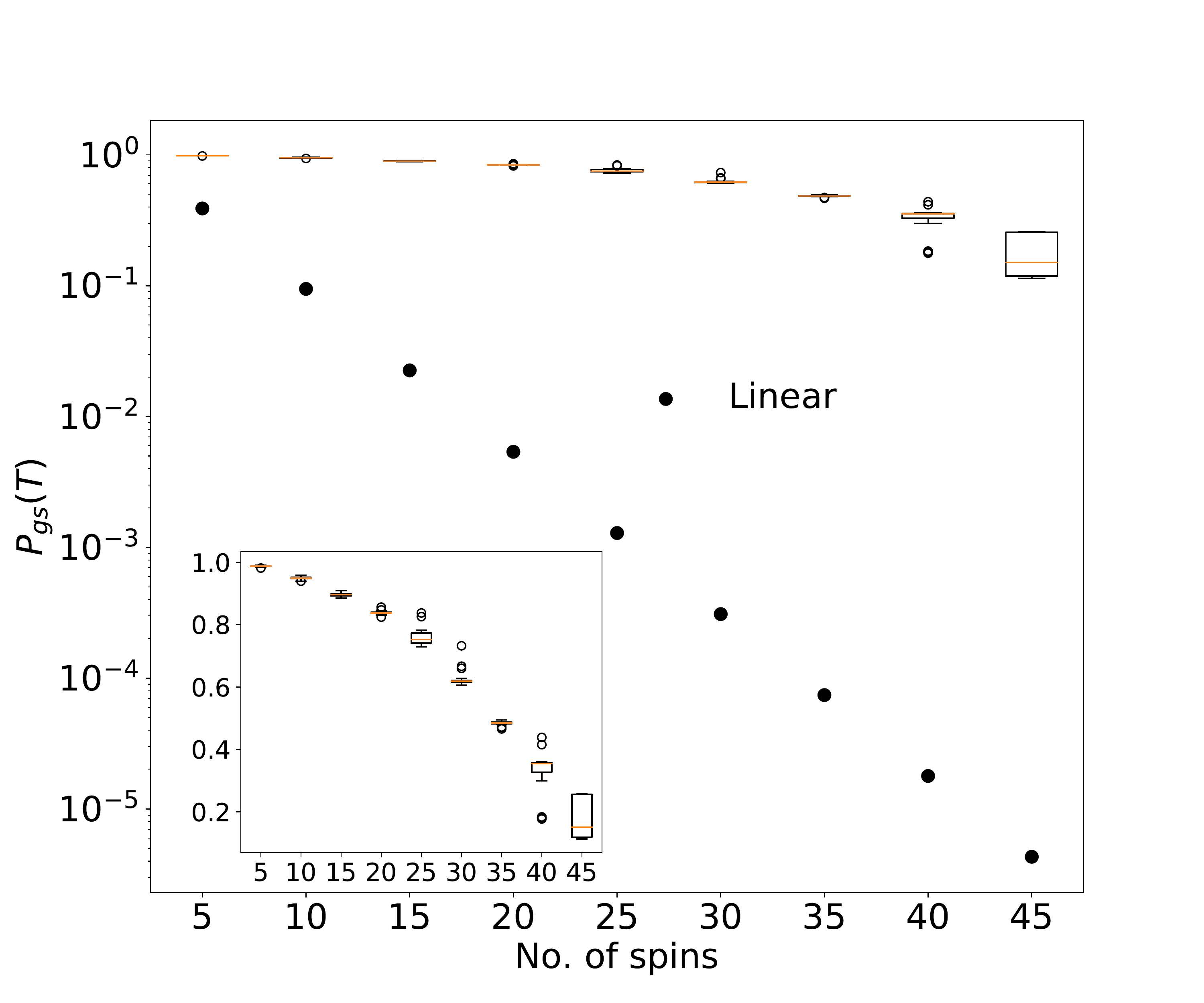}
        \caption{Boxplot of fidelities of the states of systems with different sizes. Each box represents the first quartile and the third quartile and the red line represents the median of the data for 50 runs of SOGA which optimizes the annealing schedules $A(s)$ and $B(s)$, each of which is expanded up to a third-degree polynomial.}
        \label{fig:ts-nspin}
    \end{figure}
\section{Results}\label{sec:results}
     In this section we present the results obtained by performing adiabatic quantum computation of the ferromagnetic $p$-spin model assisted by genetic algorithms. In particular, we concentrate on a system with $15$ spins and $p=3$ to demonstrate our results. The adiabatic time scale of Eq.~\eqref{eq:adiabatic-timescale} for this system is $T_\text{AD}\approx 30$.  We choose the annealing time $\tf= T_{\text{AD}} / 10\approx 3$ in order to be far from adiabaticity. 
     Throughout the time evolution, we store the data of energy gaps between the ground state and the first excited state, time schedule function values and ground state probabilities. We initiate the genetic algorithm with a population of $ \npop = 20$ individuals, and run it for a large enough number of generations until the algorithm gives convergent values. When implementing genetic algorithms, it is advisable to perform an initial experimentation to optimize the hyperparameters involved in mutation, crossover and selection processes. The details of this procedure are given in Appendix~\ref{sec:hparameters}. We consider the optimal hyperparameters to repeatedly perform genetic algorithms and to analyse the results obtained from their solutions. With the optimized annealing schedules and an optimal driving (OD) operator, the Schr\"odinger equation is solved in the time domain [0, $\tf$] and sampled at 100 evenly spaced points in this interval. The system is initialized in the ground state of $\hamx$. When we optimize the annealing schedules, we evolve the Schr\"odinger equation with the Hamiltonian in Eq.~\eqref{eq:var-total-hamiltonian}, but without the optimal driving $\hamcd(s, \epsilon)$ term. In the case of optimal driving optimization, we evolve the Schrodinger equation with the Hamiltonian in Eq.~\eqref{eq:var-total-hamiltonian}. The ground state probability of the system is computed along the genetically optimized path of quantum annealing. The Schr\"odinger equation evolution is simulated using the QuTiP library~\cite{qutip1, qutip2}. Further, we repeat the genetic algorithms 50 times and compute the corresponding results pertaining the dynamics of the system. Hereafter, we present the results obtained by using the three strategies assisted by SOGA. We discuss the cases where MOGAs can be opted over SOGAs in order to obtain meaningful results. Further, we test our methods with systems of varying number of spins.
    
    \begin{figure*}
            \centering
            \includegraphics[width=0.9\textwidth]{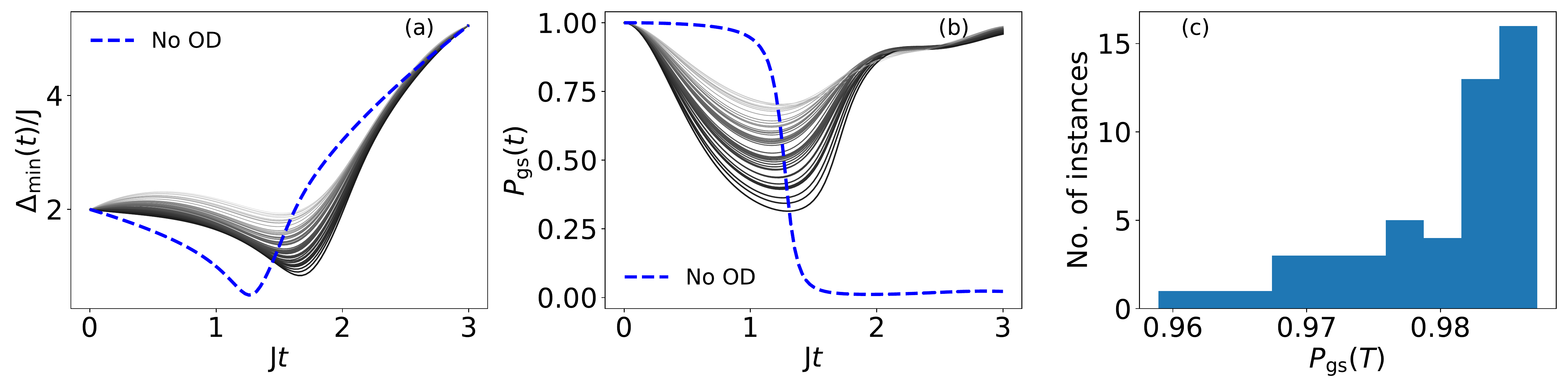}
            \caption{Results obtained by optimizing the time independent part of the local OD operator for the ferromagnetic $p$-spin model with 15 spins and $p=3$. The OD operator chosen is $H_\text{od}(s, \gamma)_{d=3}$. The plots depict the data for 50 runs of the SOGA and the corresponding results obtained by adiabatic quantum computation. a) Instantaneous energy gaps between the ground state and the first excited state. b) Instantaneous ground state probabilities. c) Histogram of the fidelities for 50 instances.}
            \label{fig:cd-so}
    \end{figure*} 
    
    \begin{figure}
            \centering
            \includegraphics[width=\columnwidth]{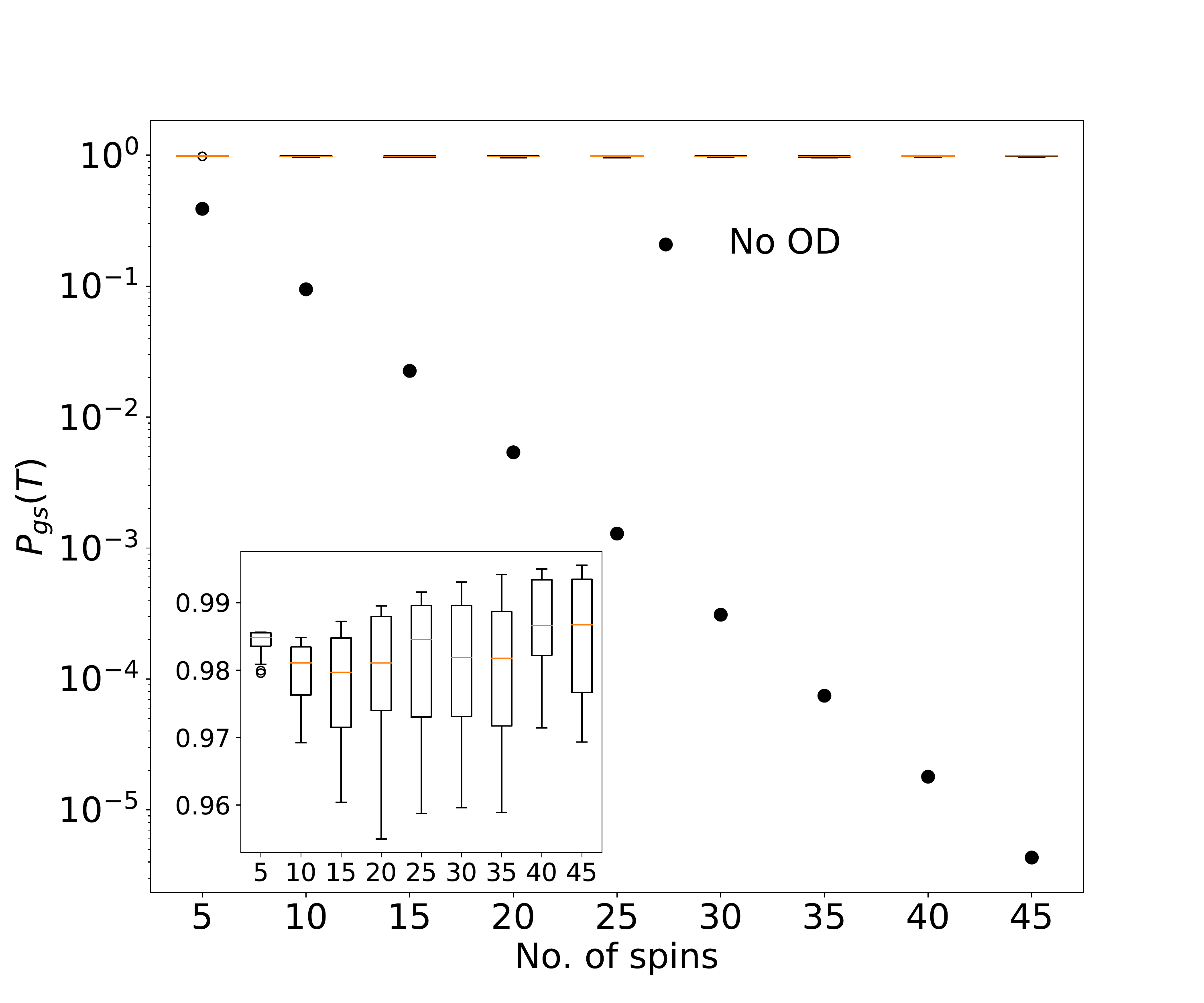}
            \caption{Boxplot of fidelities of the states of systems with different sizes. Each box represents the data for 50 runs of SOGA iterated for 1000 generations, which optimizes the time independent part of the OD operator with $d = 3$.}
            \label{fig:cd-nspin}
    \end{figure}
        
%        \begin{figure}
%        \centering
%        \includegraphics[width=0.5\textwidth]{fig-cd/chrom_contribution_all.pdf}
%        \caption{Absolute values of the chromosomes are used to visualize the contribution of each of the local operators $S_x$, $S_y$, $S_z$, in the CD Hamiltonian. The comparison has been made for different system sizes.}
%        \label{fig:chrom-contribution}
%    \end{figure}  
    
    \begin{figure}
         \centering
            \includegraphics[width=\columnwidth]{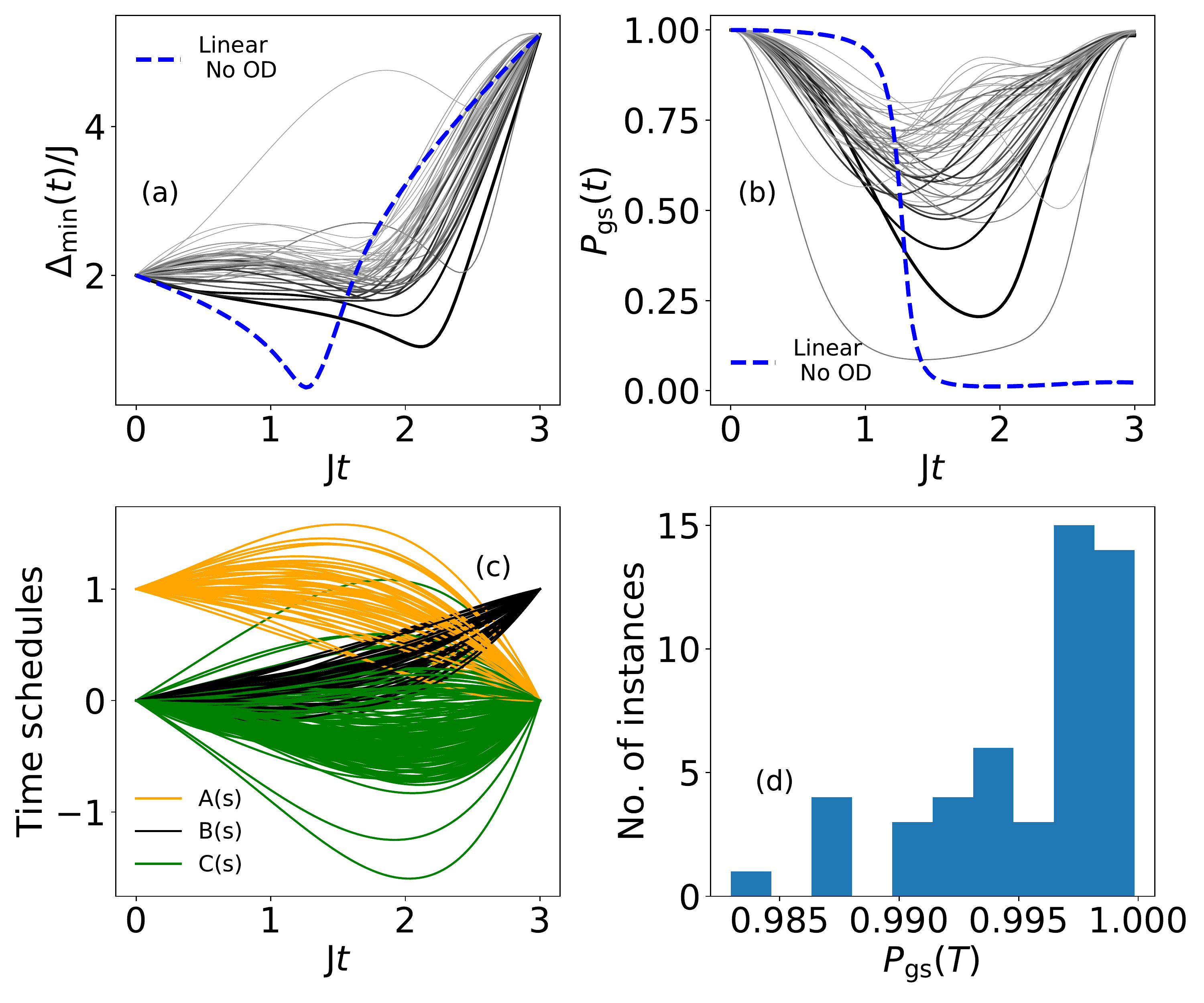}
            \caption{Results obtained by optimizing the annealing schedules $A(s)$, $B(s)$ and time dependent local OD operator for the ferromagnetic $p$-spin model with 15 spins and $p=3$. The OD operator chosen is $H_\text{od}(s, \gamma)_{d=3}$, and $k_a=k_b=2$ and $k_c=3$. The plots depict the data for 50 runs of the SOGA and the corresponding results obtained by adiabatic evolution. a) Instantaneous energy gaps between the ground state and the first excited state. b) Instantaneous ground state probabilities. c) Time schedule functions, $A(s)$, $B(s)$ and $C(s)$ d) Histogram of the fidelities for 50 instances.} 
            \label{fig:cd-ts-so}
        \end{figure}
        \begin{figure}
            \centering
            \includegraphics[width=\columnwidth]{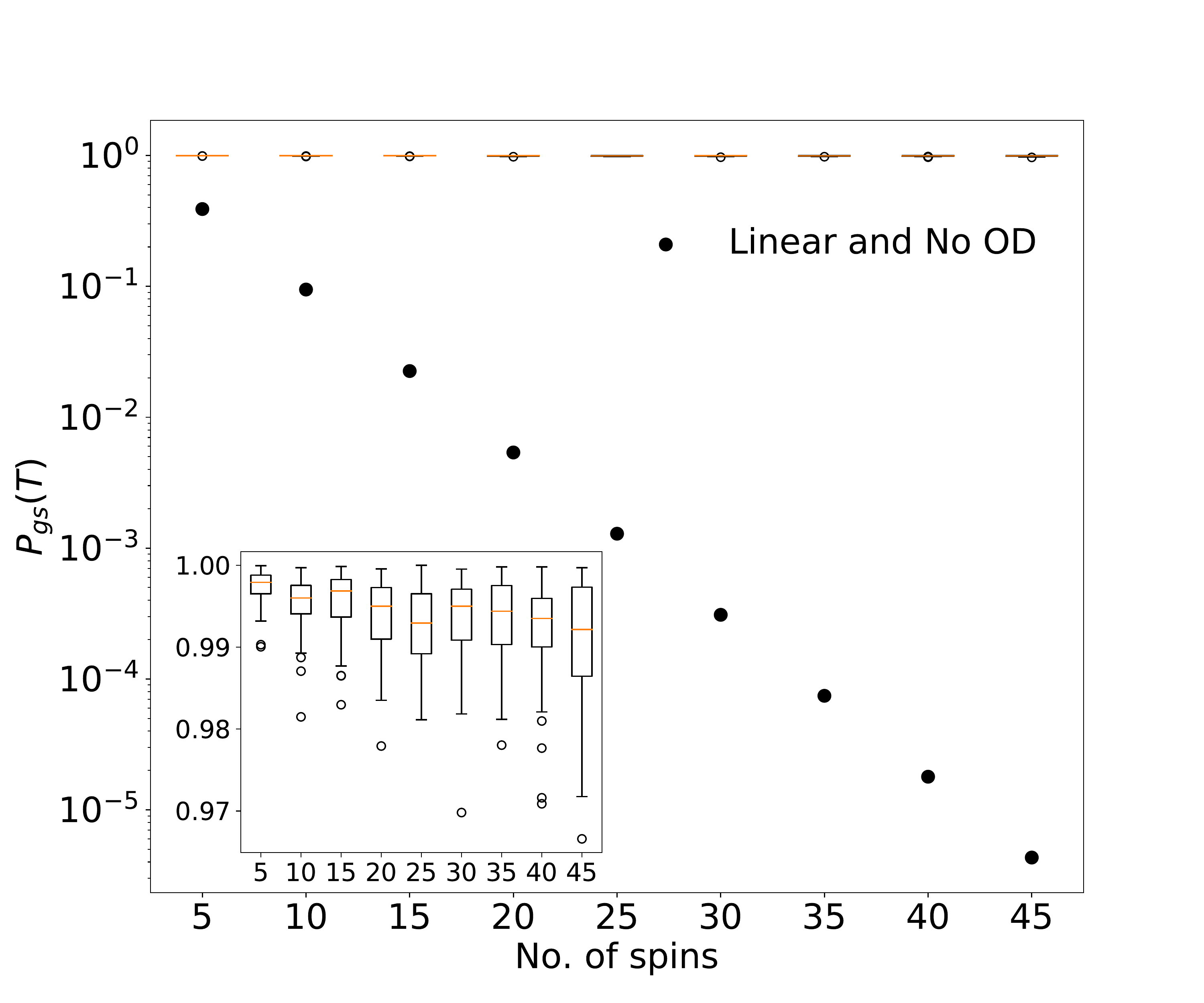}
            \caption{Boxplot of fidelities of the states of systems with different sizes. Each box represents the data for 50 runs of SOGA iterated for 1000 generations which optimizes the annealing schedules $A(s)$ and $B(s)$, and the scheduling of local OD operator [$H_\text{od}(s, \gamma)_{d=3}$], with $k_a=k_b=2$ and $k_c=3$.}
            \label{fig:cd-ts-nspin}
        \end{figure}
    \subsection{Optimization of annealing schedules \texorpdfstring{$A(s)$}{A(s)} and \texorpdfstring{$B(s)$}{B(s)}}
    As described in Sec.~\ref{sec:methods}, we optimize the annealing schedules by encoding the coefficients of the polynomials in Eq.~\eqref{eq:annealing-sch}  as chromosomes $D_1$ in Eq.~\eqref{eq:chrom-annealing-sch}. With the optimal annealing schedules given by the genetic algorithms, we simulate the adiabatic quantum computation. We focus on the cases when $k_a=2$ and $k_b=2$ and hence the length of the chromosome is 4. We have run the algorithm for 5000 generations.  
    
    The summary of the results by optimizing the annealing schedules using SOGA is provided in  Fig.~\ref{fig:annealsch-so}.
    In Fig.~\ref{fig:annealsch-so}(a), we show that the optimized path increases the minimum gap between the ground state and the first excited state only slightly: energy scales remain within practical limits. Meanwhile, in Fig.~\ref{fig:annealsch-so}(b), the ground state probabilities remain higher throughout the evolution and, around the final time, there is a slight drop in the fidelity. This could possibly be overcome by using MOGA by imposing a condition in the fitness function that the derivative of the ground state probability evolution curve remains smaller. In Fig.~\ref{fig:annealsch-so}(c), we show the annealing schedules optimized by the SOGA. We see that both the schedules $A(s)$ and $B(s)$ increase to a value larger than one and gradually decrease to their respective boundary values, as opposed to the traditionally used monotonic functions~\cite{lidar:adiabatic-theorem, acampora:genetic}. We point out that the nature of our optimized time schedules is different from the exact solution of annealing schedule function derived for example in Ref.~\cite{Roland:time-sch}. This is due to the fact that we do not impose the local adiabaticity at all points of time, but only at the final time. Fig.~\ref{fig:annealsch-so}(d) shows the histogram of fidelities for 50 runs of the algorithm. Fidelities are distributed in a small window with median value of the distribution being $\approx0.895$, which is about two orders of magnitude higher with respect to the linear schedule.

    To conclude this section, we study the genetic optimization of annealing schedules for varying system sizes. In Fig.~\ref{fig:ts-nspin}, we fix the chromosome length to be 4, and we run the genetic algorithms for 5000 generations for system sizes up to 45 spins. We plot the fidelities of the adiabatic evolution as a box plot where each box represents the interval between the first and third quartiles and the red line is the median fidelity over 50 repetitions. The solutions by the genetic algorithms decrease for larger system sizes. However, the performance is strikingly better than the corresponding results using linear annealing schedules, by several orders of magnitude.

    \subsection{Optimization of OD}
        Here we optimize the local OD operators alone fixing linear annealing schedules as described in Sec.~\ref{sec:methods}. The chromosome is $D_2$ in Eq.~\eqref{eq:chrom-local-cd-op}. We focus on optimizing the local operators with only single spin operators  $\hamcd(s, \gamma)_{d=3}$ from Eq.~\eqref{eq:cd-operators} and show that optimizing only 3 parameters, we obtain good fidelities. The higher number of local terms lead to many trivial solutions of simply increasing the energy scale of the system beyond practical capabilities due to the large solution space, at the same time being computationally expensive. 
        
        The summary of the results obtained by genetic optimization of $\hamcd(s, \gamma)_{d=3}$ are shown in Fig.~\ref{fig:cd-so}. In Fig.~\ref{fig:cd-so}(a) we show the energy gaps between the ground state and the first excited state. The minimum energy gap is slightly higher than the original system driven with no OD potentials. In panel (b), we show the corresponding results of the evolution of ground state probabilities. Even though the ground state probabilities are comparatively lower during the evolution, the fidelities are high at the final time. The probabilities can be controlled to be higher also during the evolution using MOGA. The results are not very diverse due to the small chromosome size and yet these set of solutions are feasible. Finally in Fig.~\ref{fig:cd-so}, we show the distribution of fidelities for 50 solutions of SOGA. All the solutions show very high fidelity with the median value of $\approx0.98$. We analyzed the data of optimized chromosomes to understand the contribution of each of the local operator term in the expansion of optimized local OD operator. The contribution of $S_y$ is larger for all the cases considered, which is expected since the $S_y$ term is the starting point for many known expansions of the OD operator~\cite{polkovnikov:nested-cd, polkovnikov:local-cd, passarelli:PRR2020, hartmann2019:local-cd}. 
        
        We verify the robustness of the genetic optimization approach in OD driving for larger system sizes. In Fig.~\ref{fig:cd-nspin} we compare the fidelities of states of the systems up to 45 spins. The fidelities are very high despite increasing the number of spins by optimizing only single spin operators (\ie, $d=3$). Nevertheless, when we increase the size of the system, some of the solutions given by the genetic algorithms lead to energy level crossings between the ground state and the first excited state. The corresponding ground state probabilities fall to very low values in these points and regain better values towards the end of the evolution. However, this is an unphysical scenario. We resort to MOGA in this case, which makes sure the ground state probabilities are higher throughout the evolution by avoiding the situations of energy crossings. An example of improvement of the results using MOGA for a system with 40 spins is demonstrated in Appendix~\ref{sec:pof}.

    \subsection{Optimization of time schedules and the local OD}
        Here we optimize the free parameters of the time schedules $A(s, \alpha)$, $B(s, \beta)$, and $C(s, \epsilon)$ all together as a chromosome $D_3$ in Eq.~\eqref{eq:chrom-cd-ts}. We choose $k_a=2$, $k_b=2$ and the number of local operators $d=3$, each accompanied by a time schedule $C_i(s, \epsilon)$ as described in Eq.~\eqref{eq:cd-ts} with $k_c=3$. It is sufficient to run the algorithm up to 1000 generations in this case in order to obtain convergent results.
        
        Fig.~\ref{fig:cd-ts-so} shows the summary of the results obtained by optimizing all the time schedules in the realm of shortcuts to adiabaticity. In Fig.~\ref{fig:cd-ts-so}(a), we show the minimum energy gaps. In this case, the solutions are quite diverse because of the larger search space. The same is reflected in the evolution of ground state probabilities in Fig.~\ref{fig:cd-ts-so}(b). In Fig.~\ref{fig:cd-ts-so}(c), we show the optimized annealing schedules. While some of the solutions show the same increase and decrease patterns seen in the previous case, some others are monotonic between the boundary values. The schedules $C(s)$ plotted in green color are composed of the three time functions $\{C_1(s), C_2(s), C_3(s)\}$ of each of the local operators in the expansion of the OD potential. We show the distribution of fidelities in the solutions given by the genetic algorithm in Fig.~\ref{fig:cd-ts-so}(d). The fidelities are exceptionally higher with a median value of $\approx 0.997$.
        
        In Fig.~\ref{fig:cd-ts-nspin}, we compare the fidelities of adiabatic quantum computation assisted by genetic algorithms for varying system sizes. Here we have fixed the chromosome length to be $13$ and we run the algorithm for 1000 generations for all the cases. The performance of genetic optimization is consistently higher even for larger system sizes.
        
    \section{Generalization to random Ising models}\label{sec:ising}
\begin{figure}
         \centering
         \includegraphics[width = 4 cm]{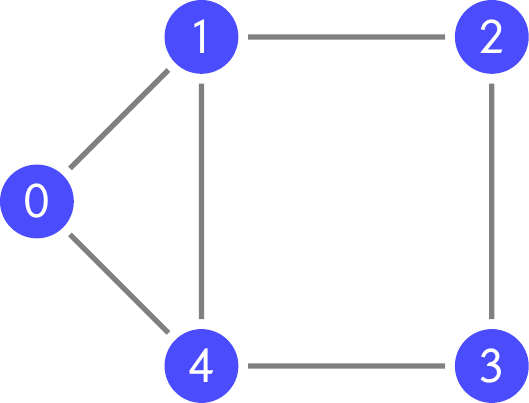}
         \caption{Graph of the Ising model discussed in Sec.~\ref{sec:ising}.}
         \label{fig:graph}
     \end{figure}

     In order to test the feasibility of our method in a more general framework, we additionally studied the performance of the genetic optimization for a random Ising model. We considered a system of $ n = 5 $ qubits arranged in the graph shown in Fig.~\ref{fig:graph}, described by the following Hamiltonian,
     \begin{equation}\label{eq:ising-model}
         H_z = H_\text{I} = \frac{1}{2}\sum_{\langle ij\rangle}\left(\mathbb{1}-J_{ij}\sigma_i^z\sigma_j^z \right),
     \end{equation}
     where the sum acts on qubits connected by the graph bonds and the couplings $J_{ij}$ are random uniform variables in $[-1, 1]$. The idea here is to apply the genetic routine to a family of randomized models so as to see if some general features of optimized annealing schedules/OD operators can be inferred. This would allow us to significantly speed up computation since it would remove the need to repeat the genetic optimization on an instance-by-instance basis. 
     \begin{figure*}
         \centering
         \includegraphics[width=\textwidth]{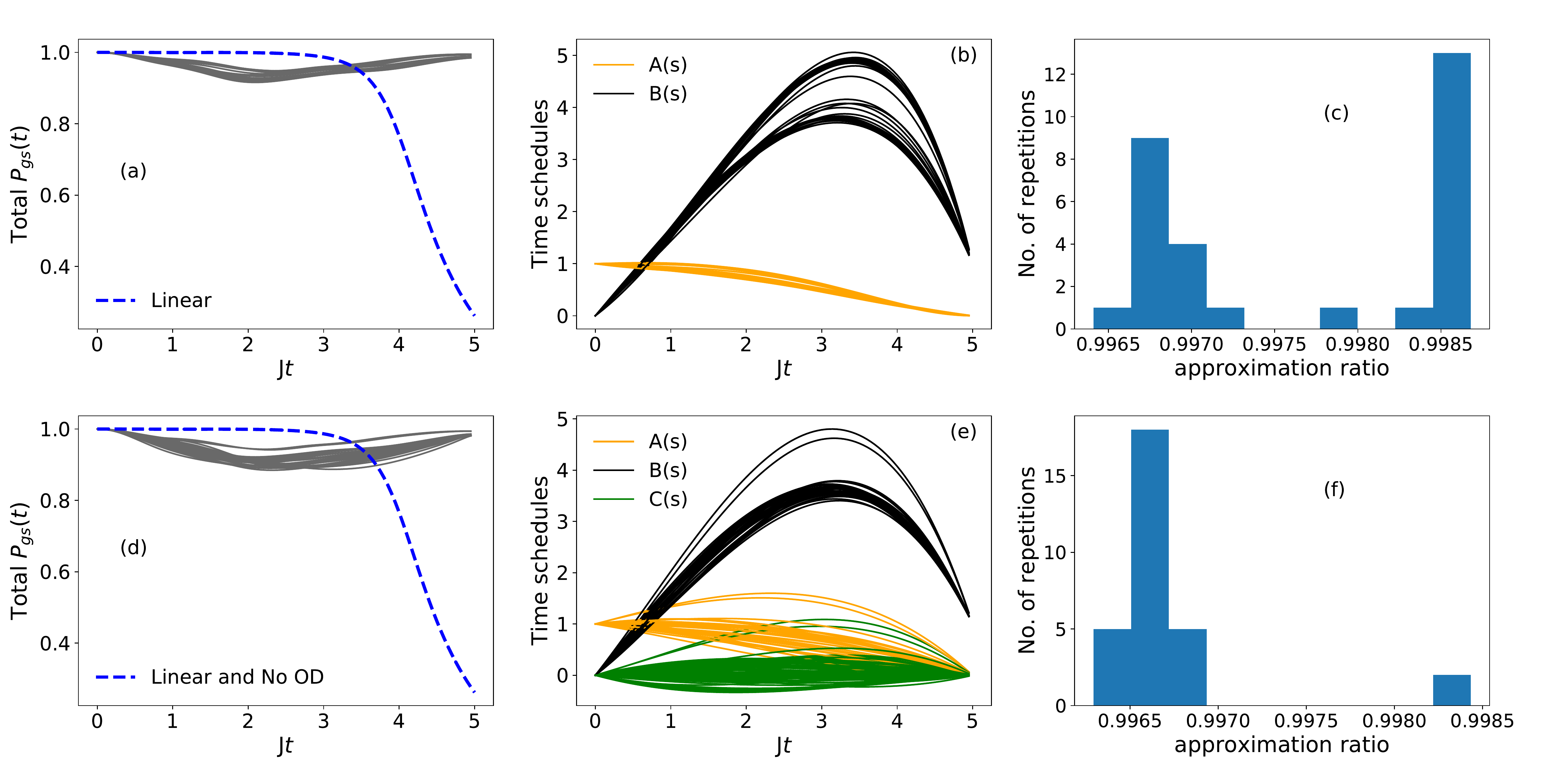}
         \caption{Results of genetic optimization of quantum annealing of a random Ising model (Eq.~\eqref{eq:ising-model}), for a typical random instance. The plot shows the data of 30 genetic optimizations of the considered random instance. Top panel depicts the optimization of annealing schedules alone, with the parameters, $k_a=k_b=3$. Here, (a) Instantaneous total probability of obtaining degenerate ground states using optimized polynomial schedules vs using linear schedules (b) Optimized annealing schedules $A(s)$ and $B(s)$ (c) Approximation ratios of 30 genetic optimizations of the given random instance. In the bottom panel, (d), (e) and (f) are the corresponding results obtained by optimizing annealing schedules and OD operator($H_\text{od}(s, \gamma)_{d=3}$) together. The parameters considered in this case are $k_a=k_b=2$ and $k_c=3$.}
         \label{fig:ising-inst1}
     \end{figure*}
     \begin{figure}
         \centering
         \includegraphics[width=0.5\textwidth]{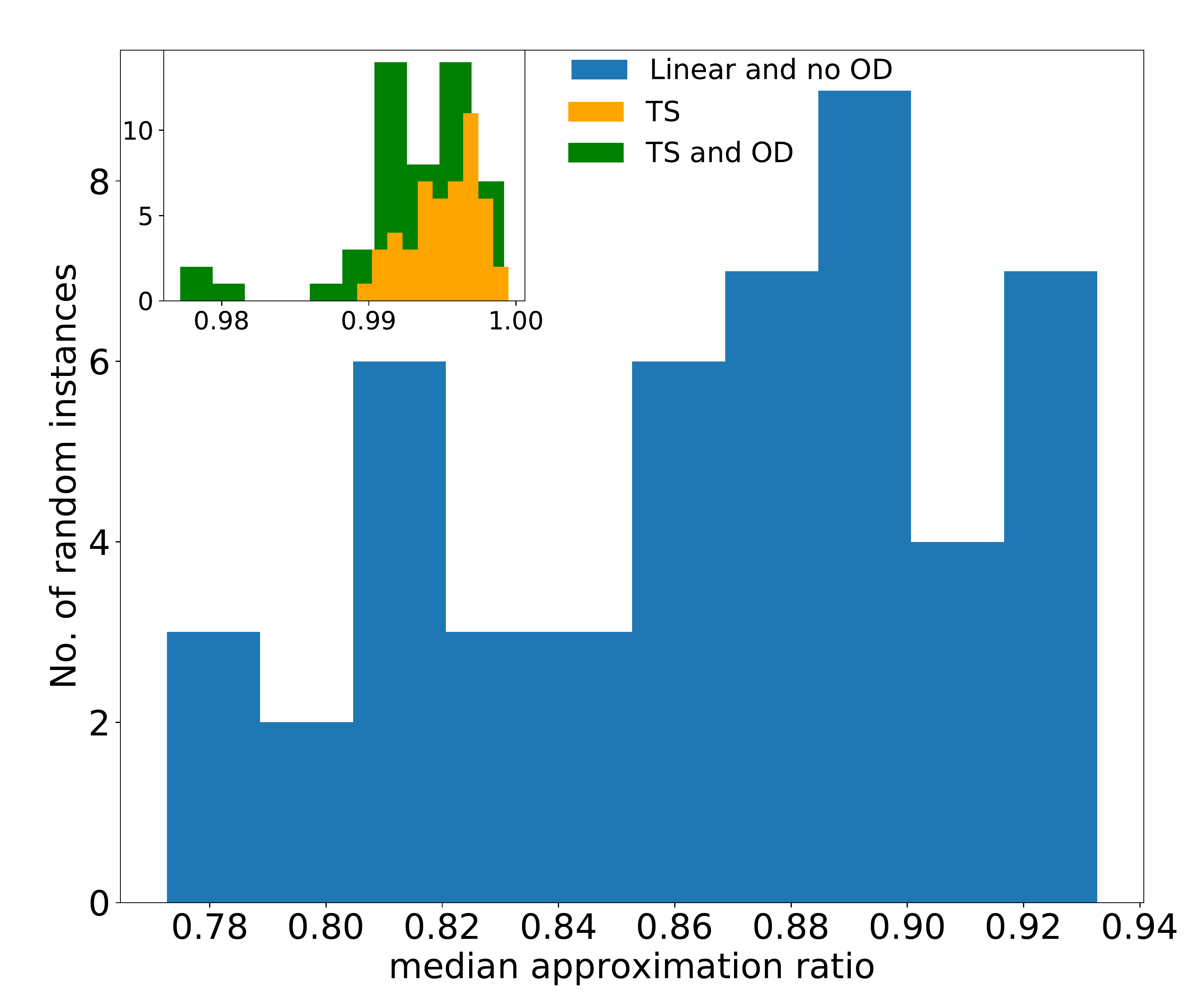}
         \caption{Median approximation ratio distribution for $N_\text{inst}=50$ random instances over $N_\text{rep}=30$ repetitions of SOGA for each instance. The histogram shows the approximation ratios of traditional quantum annealing protocol. In the inset plot, we show the corresponding approximation ratios of quantum annealing with optimized time schedules, and quantum annealing with optimized annealing schedules and OD operator.}
         \label{fig:app-ratio}
     \end{figure}
   We generated $ N_\text{inst} = 50 $ random instances and repeated the (stochastic) genetic optimization $ N_\text{rep} = 30 $ times for each instance and for each choice of the parameters of the simulation. In particular, we considered two different annealing times ($\tf = \text{\numlist{5;10}} $). For the optimization of annealing schedules alone, we considered the parameters of the polynomial ansatz, $k_a = k_b = 3$, while for the optimization of the OD operator we considered $k_a=k_b=2$ and $k_c=3$. We focused our attention on SOGAs and, since the target Hamiltonian is $Z_2$ symmetric and the ground state is doubly degenerate, we resorted to the average final energy as fitness function: $ f_\text{so} = \langle H_\text{I} \rangle $. We quickly note that the Hamiltonian in Eq.~\eqref{eq:ising-model} is commonly used to encode MaxCut and MinCut problems \cite{crooks:qaoa}. This is why, in the following, we will show data concerning the so-called approximation ratio, i.\,e. the ratio between the final fitness value and the true GS energy, which is a commonly used figure of merit in approximate optimization of this kind of problems \cite{farhi:qaoa, crooks:qaoa}.
   
   We show the results for a typical random instance in Fig. \ref{fig:ising-inst1} by optimizing annealing schedules alone and by optimizing both annealing schedules and OD operator. First, we focus on the optimization of the annealing schedules (see Fig.~\ref{fig:ising-inst1} (a)(b)(c)). In all cases analyzed, the annealing schedules are nonmonotonic like for the $p$-spin model of Sec.~\ref{sec:problem}. In addition, since the final typical energy scale is smaller than the starting one, we note that the schedule $B(s)$ is always larger than $A(s)$. The energy scales remain comparable with the ones of linear annealing schedules, but the approximation ratio is substantially improved compared with the linear schedules. The results are similar when annealing schedules and OD operators are optimized together (see Fig.~\ref{fig:ising-inst1} (d)(e)(f)). Especially, the annealing schedules continue  to show nonmonotonic features, and we note that the schedules of the OD operator $C(s)$ is bounded within a smaller range of values. Also, the approximation ratios are significantly higher than the bare case. In Fig. ~\ref{fig:app-ratio}, we compare the median approximation ratios (median of $N_{rep}=30$ SOGA repetitions) of 50 random instances of the Ising model. It is evident that the genetically optimized quantum annealing protocols show consistently higher approximation ratios than the traditional quantum annealing with linear annealing schedules and without OD.
    
    Even though the preliminary analysis of this problem shows considerably promising results, the question of whether one can find a general optimal schedules or OD operator which optimizes any random instance of Ising model remains open. We considered the average time schedule obtained from the data of 50 random instances, and investigated if this averaged schedule optimizes new random instances. In most cases it shows slight improvement when compared to the bare case. However, our analysis is far from being comprehensive in this test case and we reserve the possibility of expanding on this aspect in future works together with machine learning techniques.

    \section{Conclusions}\label{sec:conclusions}
     In this paper, we used genetic algorithms to optimize the performance of quantum annealing. We demonstrated the efficiency of our method for the ferromagnetic $p$-spin model with $p=3$. In the beginning, we optimized the annealing schedules of the standard adiabatic quantum computation protocol using genetic algorithms. We considered the time schedules to be polynomial expansions, whose coefficients were optimized as chromosomes of genetic algorithms. For a system with 15 spins, we were able to achieve a median fidelity of $\approx 0.895$, by optimizing 4 free parameters of the polynomials.
     
     We used the genetic algorithms in the paradigm of shortcuts to adiabaticity as well. Here, we optimized a practically implementable local Hamiltonian composed of only single spin operators which when added to the system Hamiltonian can improve the fidelity of the state of the system. In the first step, we fixed the annealing schedules to be linear functions of time and the time schedule of the optimal driving operator to be a quadratic function. By optimizing only the coefficients of single spin operators, \ie, by optimizing only 3 free parameters, we were able to achieve a median fidelity of $\approx 0.98$, for a system with 15 spins. As a next step, we optimized the annealing schedules, and the time-dependent coefficients of the local operators together. In this case, the time schedule of each of the optimal driving operator were absorbed as their coefficients and were assumed to be polynomial functions of time. By optimizing 13 free parameters of polynomials, we were able to obtain median fidelity $\approx 0.997$. 
     
     Further, we tested our methodology for varying system sizes. While optimizing annealing schedules alone showed a decrease in the fidelities, optimization of optimal driving showed consistent performance even for larger systems with up to 45 spins by optimizing only local single spin operators. We also discussed the cases when the single objective genetic algorithms give unphysical solutions of energy crossings and the possibility of using multi-objective genetic algorithms to tackle this problem. 
     
     We tested the technique of SOGAs for a generic case of random Ising models. We generated 50 random instances of Ising models. We separately analyzed the results when only annealing schedules are optimized with chromosome size $6$ and as well as in the picture of optimal driving with chromosome size $13$. We compared the approximation ratios (the ratio between the energy of the final state and the energy of the true ground state) of the traditional quantum annealing with those of genetically optimized quantum annealing and demonstrated that genetic algorithms are promising tools also in optimizing quantum annealing of random Ising models.
     
     In the near future, we aim to find general optimal paths of evolution for a class of random Ising models resorting to machine learning techniques. We are also going to apply the evolutionary strategies to find shortcuts to adiabaticity for open quantum systems.
     
\begin{acknowledgments}
    
We thank Giovanni Acampora, Rosario Fazio, Giuseppe Santoro and Autilia Vitiello for useful discussions and support. Financial support and computational resources from MUR, PON “Ricerca e Innovazione 2014-2020”, under Grant No. “PIR01\_00011 - (I.Bi.S.Co.)” are acknowledged. G.P. acknowledges support by MUR-PNIR, Grant. No. CIR01\_00011 - (I.Bi.S.Co.).

\end{acknowledgments}

\appendix

\section{{\label{sec:hparameters}}Optimizing the hyperparameters of genetic algorithms}
Genetic algorithms are characterized by hyperparameters pertaining the selection, crossover and mutation processes. To be precise, the individual undergo the process of mutation with a probability of $\pmut$, wherein the real numbers of the chromosome are altered according to a Gaussian distribution with variance $\sigma^2$ and mean $\mu$. Further each real number (gene) in the chromosome undergoes mutation with the probability $\pind$. We perform two-point crossover among the parent chromosomes where a string of values are cut and exchanged between the parents to produce two new solutions and this process occurring with a probability of $\pcx$. We choose the tournament selection process where among every $\tournsize$ individual chromosomes, we choose the best chromosome as parent for producing offspring. This cycle of generation repeats. In general, for each optimization problem it is advisable to perform an initial experimentation to fix these hyperparameters which give the best solution to the problem~\cite{acampora:genetic, acampora:genetic2, deap}. In particular, for the problem of annealing schedules optimization, we have tuned and chosen the hyperparameters values, $\tournsize=6$, $\pcx=0.75$, $\pmut=0.35$, $\pind=0.1$, $\sigma^2=0.6$, $\mu=0$. For the problem of finding the optimal driving, the best combination of hyper parameters is found to be $\tournsize=3$, $\pcx=0.3$, $\pmut=0.9$, $\pind=0.1$, $\sigma^2=1$, $\mu=0$. However, in this paper, for the optimization problems chosen, varying the hyperparameters have minimal effect on the overall quality of the solutions. For example, $\pmut=0.9$ gives the best fidelity, however decreasing $\pmut$ leads to searching in smaller search space which in turn reduces the number of solutions which simply increase the energy scaling of the system. Meanwhile, by doing so, the fidelity is not affected by a great deal. 
\begin{figure}
     \centering
        
        \subfigure[] {\includegraphics[width=\columnwidth]{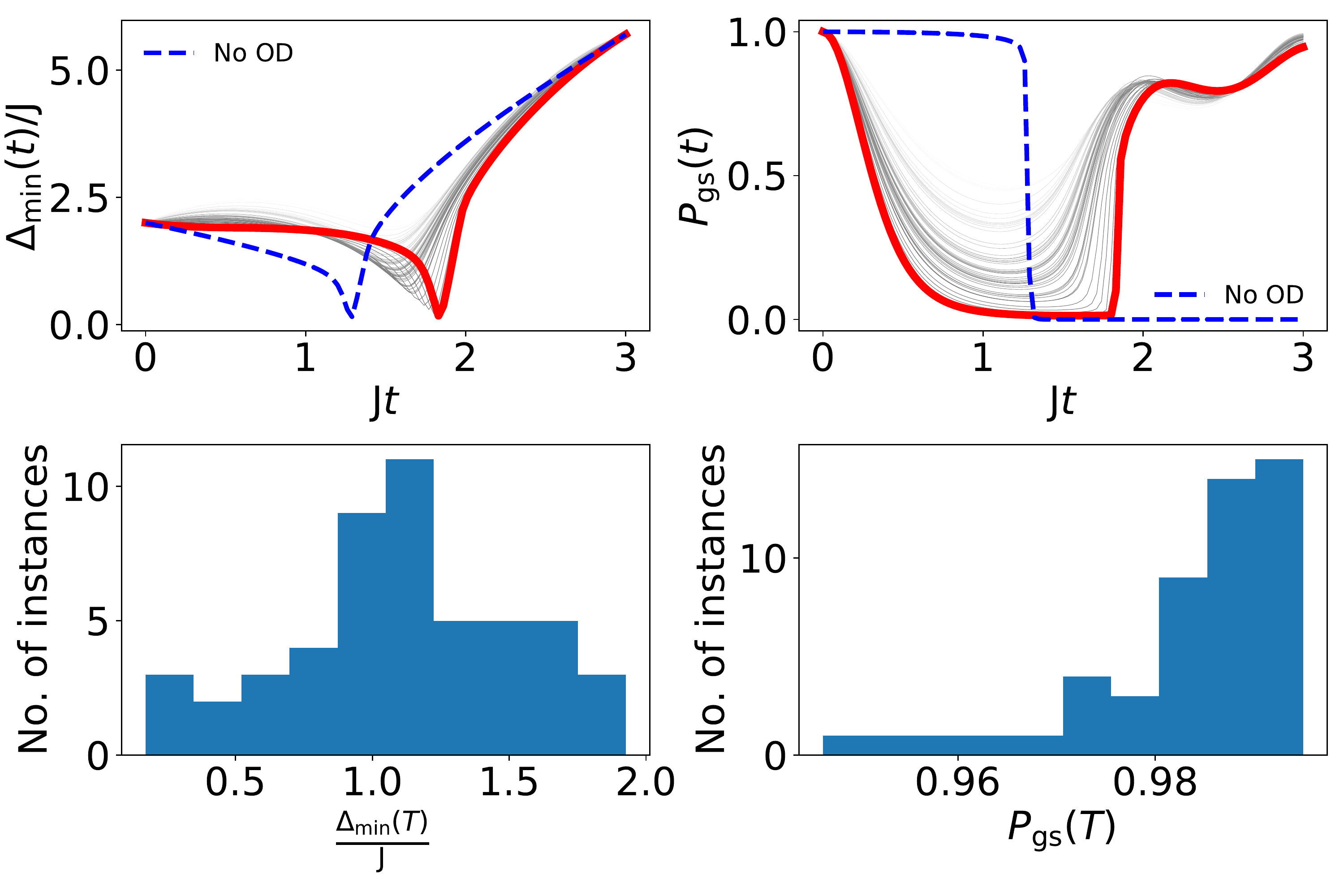}}\\
        \subfigure[]
        {\includegraphics[width=\columnwidth]{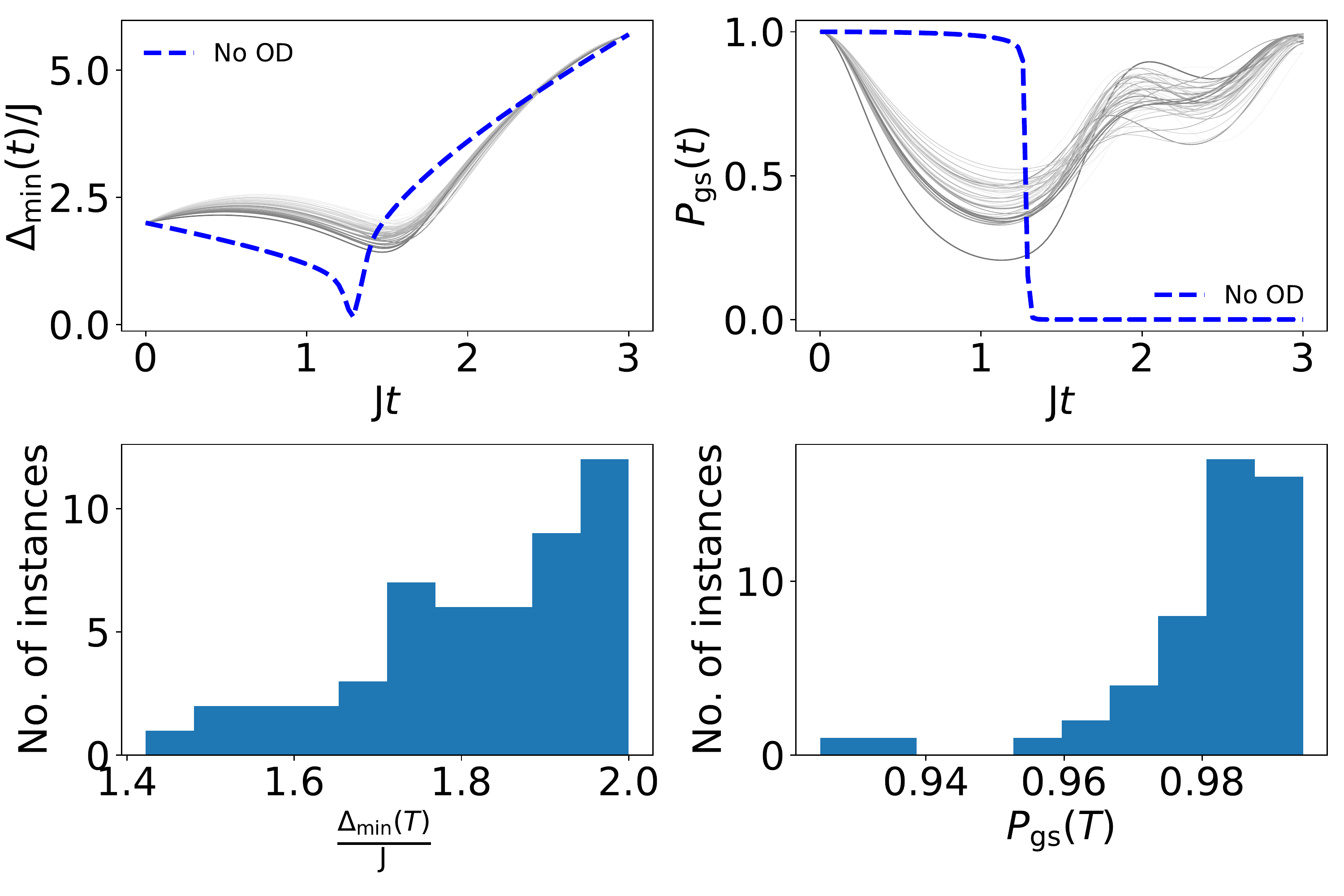}}
        
        \caption{Results of the optimization of local OD operator with 3-local operators for a $p$-spin model with 40 spins and $p=3$. We compare the performance of SOGA in (a) and MOGA in (b). The red bold line in the solutions obtained from SOGA indicate the solutions where there are energy level crossings. The corresponding results using MOGA do not show this kind of solutions.}
        \label{fig:so-vs-mo}
\end{figure}

\section{{\label{sec:pof}}Selection of chromosome from the Pareto Optimal Front in MOGA}
The output of a MOGA which is implemented using Non-dominated Sorting algorithm II, is a set of chromosomes with the best ranking in terms of their domination over the rest of the chromosomes~\cite{deb:nsga2,deap}. This set of chromosomes is called Pareto optimal front. In the end of evolution, we choose one of the chromosomes in the Pareto optimal front, which has a good trade-off between the area under the ground-state probability curve and fidelity. In this work, we choose the chromosome with the maximum value of $0.4\times\text{area} + 0.6\times\text{$P_\text{gs}(\tf)$}$ and use this solution to perform adiabatic evolution and compute results. As an example, we consider the ferromagnetic $p$-spin model with 40 spins and optimize a local optimal driving (OD) operator (with fixed annealing schedules). We show the difference in the solutions obtained from SOGAs and MOGAs in Fig.~\ref{fig:so-vs-mo}. In MOGA, with the imposition of large area under the curve of ground state probabilities, the genetic algorithm converges to solutions where there are no energy crossings. The same can be seen in the plots of $\Delta_\text{min}(t)$ and the histogram of $\Delta_\text{min}(\tf)$. The median fidelity using the results of SOGA is $\approx 0.983$, whereas with MOGA, the median fidelity is $\approx 0.981$.

%\nocite{*}

%\bibliography{genetic.bib}% Produces the bibliography via BibTeX.


\end{document}